\documentclass{an}
\usepackage{graphicx}
\usepackage{times}
\usepackage{fancyhdr}
\begin{document}

\title{Low Mass Companions to White Dwarfs}

\author{J. Farihi\inst{1,2}
\and  B. Zuckerman\inst{2}
\and  E.E. Becklin\inst{2}}

\institute{	Gemini Observatory,
		Northern Operations,
		670 North A'ohoku Place,
		Hilo, HI 96720
\and 
		Department of Physics \& Astronomy,
		University of California,
		430 Portola Plaza,
		Los Angeles, CA 90095}
		
\date{Received; accepted; published online}

\abstract{This paper summarizes the results of over
17 years of work searching for low mass stellar and
substellar companions to more than 370 nearby white
dwarfs.  Roughly 60 low mass, unevolved companions
were found and studied all together, with over 20 
discovered in the last few years, including the first
unambiguous brown dwarf companion to a white dwarf,
GD 1400B.  The resulting spectral type distributions
for companions to white dwarfs and nearby cool field
dwarfs are compared, and the implications for binary
star formation are discussed.  A brief analysis of
GD 1400B, including new data, is also presented.
}

\correspondence{jfarihi@gemini.edu}

\maketitle

\section{Introduction}

Searching for brown dwarfs as companions to stars
offers the opportunity to search systems near to Earth and
requires less time than field or cluster searches covering
a relatively large portion of the sky.  The first serious
brown dwarf candidate was discovered as a companion to the
white dwarf GD 165 (Becklin \& Zuckmerman 1988).  GD 165B
($M\sim0.072$ $M_{\odot}$, $T_{\rm{eff}}=1900$ K) remained
unique for a number of years but eventually became the
prototype for a new spectral class of cool stars and brown
dwarfs, the L dwarfs.  The first unambiguously cool brown dwarf
was also discovered as a companion to a star, Gl 229
(Nakajima et al. 1995).  Gl 229B ($M\sim0.040$ $M_{\odot}$,
$T_{\rm{eff}}=950$ K) became the prototype T dwarf, the
coolest known spectral class, all of whose members are
brown dwarfs.

The study of low mass stellar and substellar companions
to white dwarfs yields useful information regarding the
initial mass function near the bottom of the main sequence
and below, the overall binary fraction of intermediate
mass stars, the long term stability and survivability
of low mass objects in orbit about post-asymptotic giant
branch stars, and has a few advantages over similar searches
around main sequence stars (Zuckerman \& Becklin 1987; 1992;
Schultz et al. 1996; Farihi 2004; Farihi et al. 2005).

This paper summarizes results for 371 white dwarfs
which were searched for low luminosity companions using 
near-infrared imaging arrays at several facilities (mainly
Steward, Keck, \& IRTF) over the past 17+ years.  For full
details on the survey, including data acquisition, reduction
and analyses, comprehensive information on all targets,
photometry and spectra of companions as well as extensive
notes on individual objects and systems, the interested
reader is referred to Farihi (2004); Farihi et al. (2005).

\section{A Kinematically Young White Dwarf Sample}

The white dwarf targets selected for the survey were
taken almost exclusively from McCook \& Sion (1987; 1999).
In general, the selection was guided by: (1) proximity to
the Earth; (2) small to moderate proper motions; (3) youth
indicators such as mass, temperature, or cluster membership.
Selecting nearby targets has obvious sensitivity advantages
over more distant targets of a similar nature, while selecting
white dwarfs with relatively smaller proper motions aimed to
cull a sample that is not kinematically old (i.e. not thick
disk stars).

\begin{figure}[ht]
\resizebox{\hsize}{!}
{\includegraphics[]{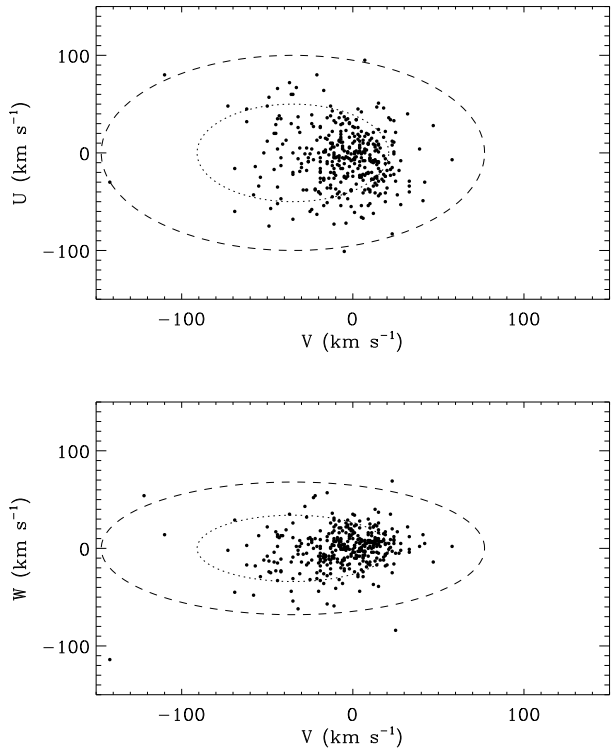}}
\caption{Galactic space velocity distribution
in the $UV$ and $WV$ planes for all 371 white dwarfs
in the sample, assuming $v_r=0$.  The ellipses represent
the 1 and 2 $\sigma$ contours for old, metal-poor disk
stars from Beers et al. (2000)}
\label{fig1}
\end{figure}

Figure \ref{fig1} displays the Galactic $UVW$ space motions
for the entire sample of white dwarfs (assuming zero radial
velocity for uniform treatment), plotted together with contours
for old, metal-poor disk stars.  The sample white dwarfs have
kinematics offset from the thick disk, and centered about zero
in $UVW$ - values that represent the undisturbed circular Galactic
disk orbits of younger stars (Mihalas \& Binney 1981; Binney \&
Merrifield 1998).

Comparing the resulting kinematical statistics of
the white dwarf sample with values for stellar populations
of known ages, from $Hipparcos$ measurements of nearby stars,
yields additional evidence that the white dwarf sample contains
young disk stars (see Farihi 2004; Farihi et al. 2005).  The average
$UVW$, their dispersions, and the total velocity dispersion values
of the entire sample are consistent with those of disk stars of
intermediate age ($\tau=2-5$ Gyr), but inconsistent with stars
of age $\tau=5$ Gyr due to the relatively small negative value
of $\langle V \rangle$ (Wielan 1974; Jahrei\ss \ \& Wielen 1997).
In fact, a subsample of 330 white dwarfs with $\mu<0.50'' \ 
\rm{yr}^{-1}$, is quite consistent with stars of age $\tau\sim2$
Gyr (Wielan 1974; Jahrei\ss \ \& Wielen 1997).

Are the cooling ages of the sample white dwarfs consistent
with a relatively young disk population?  Is the $2-5$ Gyr total
age range estimate significantly greater than the typical sample
white dwarf's cooling age?  Exactly 90\% of the 
sample stars have temperatures above 8000 K -- implying cooling 
ages less than 1.1 Gyr for typical hydrogen atmosphere white
dwarfs (Bergeron et al. 1995).  Moreover, 67\% of the sample
stars have temperatures above 11,500 K and hence typical cooling
ages less than 0.4 Gyr.  Therefore the cooling ages of the sample
stars are consistent with the total age estimate inferred from
kinematics -- that of a relatively young disk population.

Since one does not know the main sequence progenitor ages
for the white dwarf sample, caution must be taken not to
over interpret the kinematical results.  In principle, any
individual star of any age can have any velocity.  While the
sample white dwarf cooling ages are consistent with young disk
objects, a conservative approach would be to explore a range
of ages when interpreting the implications of the survey results.
Realistically, a typical white dwarf in the sample is likely
to be between $\tau=2-5$ Gyr old.

\section{Initial Mass Function for Companions}

\begin{figure}[ht]
\resizebox{\hsize}{!}
{\includegraphics[]{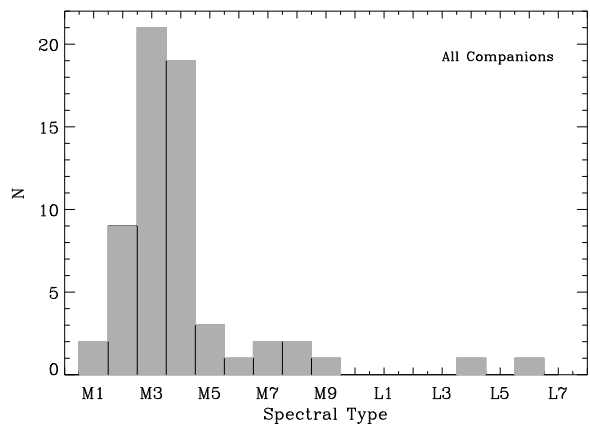}}
\caption{The number of cool dwarf companions versus
spectral type for objects discovered and studied in
the search.}
\label{fig2}
\end{figure}

The completeness limits are listed in Table \ref{tbl1}
for a typical sample white dwarf at the average distance
of 57 pc and a total age of 3 Gyr.  The sensitivity was
often greater than these conservative limits, especially
at closer distances and for younger ages.  In Figure \ref{fig2}
is plotted the number of unevolved low mass companions versus
spectral type for objects studied in this work.  Despite
excellent sensitivity to late M dwarfs and early L dwarfs
at all telescopes, very few were detected.  Additionally,
both M \& L dwarfs were detectable at arbitrarily close
separations as excess near-infrared emission (Farihi 2004,
Farihi et al. 2005), whereas T dwarfs were only detectable
as resolved, wide companions. 

\begin{figure}[ht]
\resizebox{\hsize}{!}
{\includegraphics[]{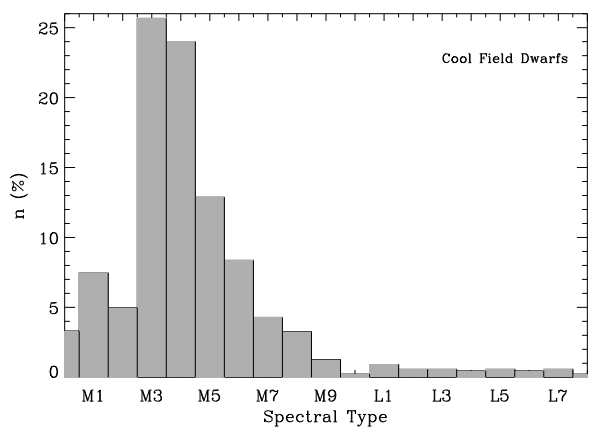}}
\caption{The frequency of cool field dwarfs within 
$d=20$ pc versus spectral type (Reid \& Hawley 2000;
Cruz et al. 2003).  The data have been corrected for
volume, sky coverage, and estimated completeness.}
\label{fig3}
\end{figure}

For comparison, Figure \ref{fig3} shows similar
statistics for cool field dwarfs within 20 pc of
Earth taken from Reid \& Hawley (2000); Cruz et al.
(2003).  The data plotted in Figure \ref{fig3} have
been corrected for volume, sky coverage, and estimated
completeness.  Can one reconcile Figure \ref{fig3} with
the common notion that there are at least as many brown
dwarfs as low mass stars (Reid et al. 1999)?  To resolve
this possible discrepancy, most field brown dwarfs would
have to be of spectral type T or later, since it is
clear from the figure that, in the field, L dwarfs
are much less common than stars.

However, there are several things to keep in mind
regarding the relative number of field brown dwarfs
versus stars.  There should be be a relative dearth
of L dwarfs compared to T type and cooler brown dwarfs
in the field because cooling brown dwarfs pass through
the L dwarf stage relatively rapidly.  The lower end of
the substellar mass function is poorly constrained at 
present (Burgasser 2004) and the relative number of
substellar objects versus low mass stars in the field
depends on the shape of the mass function in addition
to the unknown minimum mass for self-gravitating substellar
objects (Reid et al. 1999; Burgasser 2004).  Furthermore,
even for only moderately rising mass functions, such as
those measured for substellar objects in open clusters
(Hillenbrand \& Carpernter 2000; Luhman et al. 2000; Hambly
et al. 1999; Bouvier et al. 1998), there will be more brown
dwarfs than stars if the minimum self-gravitating substellar
mass is $<0.010$ $M_{\odot}$.  Ongoing and future measurements
of the local T dwarf space density will constrain the substellar
field mass function.

\begin{table}[ht]
\caption{Survey Completeness for $d=57$ pc, $\tau=3$ Gyr}
\label{tbl1}
\begin{tabular}{ccccccc}
\hline

Survey	&$a_{in}$&$a_{out}$	&$m_{abs}$	&SpT	&$M$		&N\\
	&(AU)	&(AU)		&(mag)		&	&($M_{\odot}$)	&\\

\hline

IRTF		&0	&700	&$M_K=12.2$	&L6	&0.065	&82\\
Steward		&110	&4700	&$M_J=14.2$	&L7	&0.060	&261\\
Keck		&55	&1100	&$M_J=17.2$	&T9$^{\dag}$
							&0.030	&86\\
All		&0	&110	&$M_H=13.5$	&L8	&0.058	&371\\

\hline
\end{tabular}

\medskip

$^{\dag}$ No objects are known with spectral type later
than T8.  However, the average limiting magnitude of the
Keck survey probed $\sim1.5$ magnitudes deeper than that
of any known brown dwarf (Vrba et al. 2004; Legget et al.
2002).

\medskip

This table presents only average separations and
sensitivities.  The actual values depend on each 
individual white dwarf distance and age.  The ``All''
entry refers to detection in 2MASS of an $H$ band
excess above that expected from the white dwarf
photosphere (see Farihi 2004; Farihi et al 2005).

\end{table}

Figures \ref{fig2} \& \ref{fig3} are quite similar.
Clearly, the peak frequency in spectral type occurs
around M3.5 for both field dwarfs and companions to
white dwarfs.  In fact, the peak is identical; 25.6\% 
for both populations.  By itself, this could imply a
common formation mechanism, a companion mass function
similar to the field mass function in this mass range,
approximately $0.15-0.60$ $M_{\odot}$ for spectral types
M0$-$M5 (Farihi 2004, Farihi et al. 2005).  But, relative
to the peak, there are $\sim2-3$ times more L dwarfs and 
$\sim4-5$ times more M6$-$M9 dwarfs in the field than
companions.  For the T dwarf regime, uncertainty remains
because only the Keck portion of the white dwarf survey
was sensitive to such cool brown dwarfs (and only for
certain separations) plus the current incomplete
determination of the field population density.

Hence, binary systems with small mass ratios 
$(q=M_2/M_1<0.05)$ are rare for white dwarf
progenitors (which typically have main sequence
masses $\sim2$ $M_{\odot}$).  Although there exists
some speculation regarding the possibility that brown
dwarfs are ejected in the early stages of multiple
system or cluster formation, there is currently no
evidence of this occurring.  It is conceivable that
low mass companions in very wide orbits may be lost
to gravitational encounters in the Galactic disk
over a few billion years, but given the fact that
there are a dozen or so known L and T dwarfs in wide
binaries, this seems like a rare mechanism, if it 
occurs at all.

In a way, the relative dearth of late M dwarfs alleviates a
potential interpretation problem.  Had it been the case that
many late M dwarfs were detected but only one or two L dwarfs,
it might have been argued that the L dwarfs were cooling beyond
the sensitivity of the search.  Since all M dwarfs (and the first
few L dwarf subclasses) at $\tau\geq1$ Gyr are stellar according
to theory, this concern does not exist.  The measured dearth
is real and is not caused by brown dwarf cooling and the
resulting lower sensitivity.

\section{GD 1400B}

Discovered 17 years after GD 165B (Becklin \&
Zuckerman 1988), GD 1400B is a long sought datum in
the search for low mass companions to white dwarfs
(Farihi 2004; Farihi et al. 2005).  Little is known
about the probable white dwarf plus
brown dwarf
spectroscopic binary, GD 1400.  The cool companion
was discovered,
then confirmed, through photometric
excess and subsequent
spectroscopy in the $2.2\mu$m
region (Farihi \& Christopher 2004).  Its apparent
lack of excess emission at $1.2\mu$m implies that GD
1400B has a spectral type of L5.5 or later and the
lack of Na in its $K$ band spectrum indicates it cannot
be an early L dwarf.  Utilizing the best available data
on the white dwarf primary to assess its distance and
to account for its contribution at near-infrared
wavelengths, the absolute magnitude of GD 1400B
would place it around spectral type L6 (Farihi
\& Christopher 2004).  Subsequently, an independent
spectroscopic study estimated GD 1400B at spectral
type L7 through simultaneous fits of the white dwarf
and brown dwarf components in an HK grism observation,
with model and empirical template spectra respectively
(Dobbie et al. 2005).

\begin{figure}[ht]
\resizebox{\hsize}{!}
{\includegraphics[]{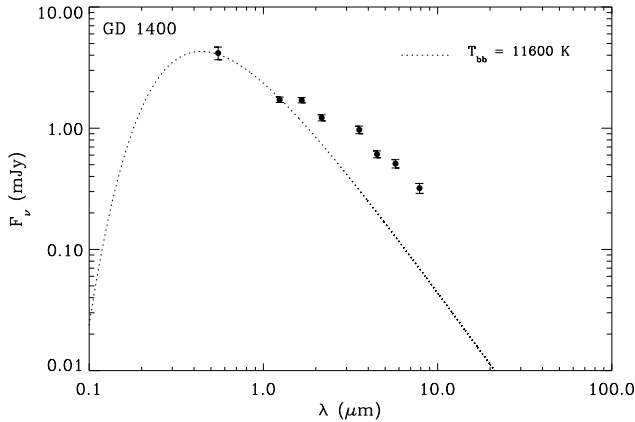}}
\caption{Spectral energy distribution of GD 1400,
demonstrating the presence of the spatially unresolved
cool brown dwarf companion.  Optical and near-infrared
data ($VJHK$) are from Farihi \& Christopher (2004)}
\label{fig4}
\end{figure}

GD 1400 has been observed with {\em Spitzer}/IRAC 
at $3-8\mu$m as part of an ongoing program
searching
for substellar companions
to nearby white dwarfs.  The IRAC
measurements of GD 1400 presented in Figure \ref{fig4} have
${\rm S/N}>15$ at all wavelengths.  The deconvolved magnitudes
of GD 1400B imply $2-8\mu$m colors consistent with a spectral
type of L5 or later, corroborating previous findings by
alternate methods (Patten et al. 2004; Farihi et al. 2005,
submitted).

Because GD 1400AB has yet to be spatially
resolved (Farihi \& Christopher 2004), it
remains possible that this spectroscopic binary
is a radial velocity variable.  It is perhaps more
likely the system resides in close orbit due to the
fact that post-asymptotic giant branch (AGB) evolution
predicts a bimodal distribution of orbital semimajor
axes for low mass, unevolved companions to white dwarfs
(Farihi 2004).  Specifically, companions close enough to
orbit within the AGB envelope should spiral inward due to
transfer of orbital energy into the envelope via friction,
while those outside the envelope should spiral outward due
to weakened gravity from mass loss (Zuckerman \& Becklin
1987; Burleigh et al. 2002; Farihi 2004).  

Further radial velocity monitoring of the white
dwarf in the optical and/or its companion in the
near-infrared, or high resolution ground- or
space-based imaging should eventually reveal
the nature of the current orbital separation
of the binary.  Resolving the pair would be
advantageous because the companion could be directly
studied.  On the other hand, it would be fortuitous
if the system were a radial velocity variable because
then the mass and radius of the secondary could be
estimated.  Currently, there is only a single L dwarf
(binary) system with a mass measurement (Bouy et al.
2004), and no mass estimates for old brown dwarfs.
There exist two independent and reliable spectroscopic
fits of $T_{\rm eff}$ and log $g$ for GD 1400A,
and hence the mass of the white dwarf is fairly
well constrainted near $M\approx0.7M_{\odot}$.
A trigonometric parallax and high precision optical
photometry would tighten up the primary mass estimate,
making any secondary mass determination more reliable.

Determining the orbital parameters of this
so far unique binary is critical to understanding
the origin and evolution of the brown dwarf secondary.
It is likely that the system formed as a extreme low
mass ratio binary ($M_2/M_1\approx0.02$; Farihi \&
Christopher 2004), but it is conceivable that the
companion formed in a massive disk around a $\sim3$
$M_{\odot}$ main sequence star.  There have been several
substellar companions detected around K giants (Frink et
al. 2002; Mitchell et al. 2003), which are the descendents
of main sequence A \& F stars.  Presumably, these substellar
companions formed in their respective primary progenitor disks
based on their current orbital semimajor axes.  Will these brown
dwarfs survive the current first ascent and ensuing asymptotic
giant branches to become companion systems similar to GD 1400?
Although complete evaporation or inspiral collision with the
stellar core is possible inside the AGB envelope, the higher
mass brown dwarfs around these K giants may persist, as has
GD 1400B, either by eschewing the greatly expanded photosphere
or simply surviving the envelope itself (Farihi et al. 2005,
submitted).

\section{Conclusions}

Together, the various phases of this survey discovered over 40 
previously unrecognized white dwarf binary and multiple systems.
The search conducted at Steward Observatory alone discovered at
least 20 new white dwarf multiple systems.  Based on the analysis
of Farihi (2004); Farihi et al. (2005) there is no reason why all
unevolved secondary stars should not be included in any initial
companion mass function, for which Figure \ref{fig2} is a good
proxy (see Farihi 2004; Farihi et al. 2005 for details).

The calculated fraction of white dwarfs with substellar
companions, within the range of masses and separations to which
this work was sensitive, is $f_{bd}=0.4\pm0.1$\%.  This represents
the first measurement of the low mass tail of the companion mass
function for intermediate mass stars, main sequence A and F stars
(plus relatively few B stars) with masses in the range 1.2 $M_{\odot}$
$<M<8$ $M_{\odot}$.  This value is consistent with similar searches 
around solar type main sequence stars for comparable sensitivities 
in mass and separation (Oppenheimer et al. 2001; McCarthy \& Zuckerman
2004).  Therefore that the process of star formation eschews the
production of binaries with $M_2/M_1<0.05$ is clear from the
relative dearth of both L and late M dwarfs discovered in
this work.

\acknowledgements

The authors thank Steward Observatory for the use
of their facilities over the years.  Part of the data
presented herein were obtained at Keck Observatory,
which is operated as a scientific partnership among
the California Institute of Technology (CIT), the
University of California  and the National Aeronautics
and Space Administration (NASA).  Some data presented
are based on observations made with the Spitzer Space
Telescope, which is operated by the Jet Propulsion
Laboratory (JPL)/CIT under NASA contract 1407.  This
publication  makes use of data acquired at the NASA
Infrared Telescope Facility, which is operated by the
University of Hawaii under Cooperative Agreement no.
NCC 5-538 with NASA, Office of Space Science, Planetary
Astronomy Program.  Some data used in this paper are part
of the Two Micron All Sky Survey, a joint project of the
University of Massachusetts and the Infrared Processing
and Analysis Center/CIT, funded by NASA and the National
Science Foundation (NSF).  The authors acknowledge the 
Space Telescope Science Institute for use of the Digitized
Sky Survey.  This research has been supported in part by
grants from NSF to UCLA and by NASA through Contract
Number 1264491 issued by JPL/CIT.

\end{document}